\documentstyle[11pt]{article}

\newcommand{\be}{\begin{equation}}
\newcommand{\ee}{\end{equation}}
\newcommand{\ba}{\begin{eqnarray}}
\newcommand{\ea}{\end{eqnarray}} 
\newcommand{\lb}[1]{\label{#1}}  
\newcommand{\bb}[1]{\bibitem{#1}}

\newcommand{\e}{{\mathrm e}}

\begin{document}
\date{October 26, 1999}
\title{Generating rotating fields in general relativity\footnote{Talk 
presented at the 10$^{th}$  Russian  
Gravitational Conference, Vladimir (Russia) 20-27 June 1999}}
\author{G{\'e}rard Cl{\'e}ment\thanks{Present address: 
Laboratoire de Physique Th\'eorique LAPTH, B.P.110, F-74941 
Annecy-le Vieux cedex, France. E-mail: gclement@lapp.in2p3.fr}\\
{\small Gravitation et Cosmologie Relativistes}\\ 
{\small Universit\'e Pierre et 
Marie Curie, Paris, France}}
\maketitle

\noindent {\bf Abstract.} I present a new method to generate rotating
solutions of the Einstein--Maxwell equations from static solutions, 
give several examples of its application, and
discuss its general properties.

\bigskip

\bigskip

When dealing with exact stationary solutions of the Einstein equations,  
one sometimes stumbles on the questions, quite easy 
to ask, but rather difficult to answer: Given some static solution, 
what is the family of non-static (rotating) solutions which are near 
---in some sense--- this static solution? On how many parameters do 
these solutions depend? And (last but not least) how can one
practically generate these rotating solutions from the static solution?

In principle, these questions can be answered in the context of the 
Geroch group \cite{geroch}. Let us recall that the 4--dimensional 
stationary Einstein (resp. Einstein--Maxwell) equations are invariant 
under the group O(2,1) (resp. SU(2,1)) of 
generalized Ehlers--Harrison transformations \cite{exact}. 
In the case of stationary axisymmetric
solutions, with two commuting Killing vectors $\partial_t$ and 
$\partial _\varphi$, the combination of the invariance
transformations associated with a given direction in the Killing 2--plane
with rotations in this plane leads to  the infinite--dimensional Geroch group.
These transformations allow in principle the
generation of all solutions of the stationary axisymmetric Einstein (or
Einstein--Maxwell) problem, which is thus completely integrable.  
This generation of stationary axisymmetric solutions can be achieved in a
variety of manners, the most popular being  inverse--scattering transform methods
applied to the Einstein (or Einstein--Maxwell) problem \cite{bz}.
While these techniques have led to the construction of many new solutions,  
until very recently no practical approach to the direct generation of rotating 
solutions from static solutions was available. 

In this talk, I shall exhibit a specific Geroch transformation which achieves 
this goal in the case of the Einstein--Maxwell theory,  using finite 
combinations of SU(2,1) transformations
and global coordinate transformations mixing the two Killing vectors
\cite{kerr, lring}. After recalling briefly the Ernst approach to the reduction
of the stationary Einstein--Maxwell problem, I shall describe this direct
rotation--generating transformation, and give several examples of its
application. I shall then discuss some general properties of the solutions 
generated by this technique, and mention  several promising lines of future research.

In the absence of matter sources, stationary Einstein--Maxwell fields 
may be parametrized by the metric 
\be\label{stat1}
ds^2 = f\,(dt - \omega_i dx^i)^2 - f^{-1}\,h_{ij}\,dx^i dx^j\,,
\ee
and the electromagnetic fields
\be\lb{stat2}
F_{i0} = v,_i\,, \qquad F^{ij} = f\,h^{-1/2}\epsilon^{ijk}u,_k\,,
\ee 
where the  fields $f$, $\omega_i$, $v$, $u$ and the reduced spatial metric
$h_{ij}$ depend only on the space coordinates $x^i$. 
The vector potential $\omega_i$ may be dualized to the scalar twist 
potential $\chi$ by
\begin{equation}\label{twist}
\chi'_i  = -f^2\,h^{-1/2}h_{ij}\,\epsilon^{jkl}\omega_l,_k 
+ 2(u\,v,_i - v\,u,_i)\,.
\end{equation}
The complex Ernst potentials are related to the four real scalar potentials 
$f$, $\chi$, $v$ and $u$  by
\begin{equation}
{\cal E} = f + i \chi - \overline{\psi}\psi\,, \qquad \psi = v + iu\,.
\end{equation}
The stationary Einstein--Maxwell equations then reduce to the
three--dimensional Ernst equations \cite{er}
\begin{eqnarray}\label{ernst2}
f\nabla^2{\cal E} & = & \nabla{\cal E} \cdot (\nabla{\cal E} + 
2\overline{\psi}\nabla\psi)\,,\nonumber \\
f\nabla^2\psi & = & \nabla\psi \cdot (\nabla{\cal E} + 
2\overline{\psi}\nabla\psi)\,, \\
f^2R_{ij}(h) & = & {\rm Re} 
\left[ \frac{1}{2}{\cal E},_{(i}\overline{{\cal E}},_{j)} 
+ 2\psi{\cal E},_{(i}\overline{\psi},_{j)}
-2{\cal E}\psi,_{(i}\overline{\psi},_{j)} \right]\,, \nonumber
\end{eqnarray}
where the scalar products and covariant Laplacian are computed with the
reduced spatial metric $h_{ij}$, These equations are invariant under an 
SU(2,1) group of transformations.

The direct rotation--generating transformation is the product
\begin{equation}\label{sig}
\Sigma = \Pi^{-1}\,{\cal R}(\Omega,\gamma)\,\Pi
\end{equation}
of three successive transformations, two ``vertical'' transformations
$\Pi\,,\Pi^{-1} \in$ SU(2,1) acting on the potential space, and a
``horizontal'' global coordinate transformation ${\cal R}(\Omega,\gamma)$ acting
on the Killing 2--plane. The transformation $\Pi$ is the SU(2,1)
involution  $({\cal E}, \psi, h_{ij}) \leftrightarrow 
(\hat{\cal E}, \hat{\psi}, \hat{h}_{ij})$ with
\begin{equation}\label{inv}
\hat{\cal E} = \frac{-1 + {\cal E} + 2 \psi}{1 - {\cal E} + 2 \psi}\,, \quad 
\hat{\psi} = \frac{1 + {\cal E}}{1 - {\cal E} + 2 \psi}\,, \quad
\hat{h}_{ij} = h_{ij}\,.
\end{equation}
The resulting transformation of the gravitational potential $f$ is 
$\hat{f} =  4f/|1-{\cal E}+2\psi|^2$.
Starting from  the Schwarzschild solution, written in prolate spheroidal coordinates
\cite{zip} as
\ba
ds^2 & = & fdt^2 - f^{-1}m^2\,[dx^2 + (x^2-1)(d\theta^2 + 
\sin^2\theta\,d\varphi^2)]\,, \nonumber \\
{\cal E} & = & f\; = \;\frac{x-1}{x+1}\,,  \qquad \psi = 0
\ea
(the coordinate $x$ is related to the ``standard'' radial coordinate $r$
by  $x = (r-m)/m$),  the action of $\Pi$ leads, after rescaling  the time
coordinate to $\tau = m^{-1}t$, and putting $y = \cos\theta$, to the 
open Bertotti--Robinson (BR) solution \cite{br}
\ba\lb{BR}
d\hat{s}^2  & = & m^2\, \left[ \,(x^2-1)\,d\tau^2 - \frac{dx^2}{x^2-1}  - 
\frac{dy^2}{1-y^2} - (1-y^2)\,d\varphi^2\, \right] \,, \nonumber \\
\hat{\cal E} & = & -1\,, \qquad \hat{\psi} = x\,.
\ea
This non--asymptotically flat spacetime is the direct product adS$_2$ 
$\times$ S$^2$
of two constant curvature two--dimensional spaces.
More generally, if the initial Ernst potentials have the asymptotic
$(r \to \infty)$ monopole behaviour $({\cal E}, \psi) \to ({\cal E}_{\infty}, \psi_{\infty}) 
+ O(1,r)$, with
\be
{\cal E}_{\infty} = 1 + 2\psi_{\infty}\,,
\ee 
the transformation $\Pi$ leads to asymptotically BR--like potentials 
$(\hat{\cal E}, \hat{\psi}) \to \;$ (const., $O(r))$ with $f \to O(r^2)$.
 
The global coordinate transformation ${\cal R}(\Omega,\gamma)$ is the
product of the transformation to a uniformly rotating frame and of a time
dilation,
\ba\label{R}
d\varphi & = & d\varphi' + \Omega\gamma\,dt'\,, \nonumber \\
dt & = & \gamma\,dt'\,.
\ea
In the case of electrostatic solutions with $\hat{{\cal E}}$ and $\hat{\psi}$
real ($\hat{\omega} = 0$), the frame rotation gives rise to an induced
gravimagnetic field $\hat{\omega}'$ as well as to an induced magnetic
field. However this transformation does not modify the leading asymptotic
behavior of the BR metric or of asymptotically BR--like metrics. Because
of this last property, the last transformation $\Pi^{-1}$ in (\ref{sig})
then leads to asymptotically flat, but complex, Ernst potentials
corresponding to a monopole--dipole solution.

For instance, the transformation (\ref{R}) leads from the BR solution 
(\ref{BR}) to 
\ba\lb{BR2}
d\hat{s}'^2  & = & m^2\, [\,\gamma^2(x^2+\eta^2y^2-(1+\eta^2))(d\tau' - 
\frac{\eta(1-y^2)}{\gamma(x^2+\eta^2y^2-(1+\eta^2))}\,d\varphi')^2 
\nonumber \\
& & - \frac{dx^2}{x^2-1}  - \frac{dy^2}{1-y^2} - 
\frac{(x^2-1)(1-y^2)}{x^2+\eta^2y^2-(1+\eta^2)}\,d\varphi'^2\, 
]\,, \\
\hat{\cal E'} & = & -\gamma^2(1+\eta^2)\,, \qquad \hat{\psi'} = 
\gamma(x-i\eta y)\,, \nonumber
\ea
with $\eta \equiv m\Omega$. The Ernst potentials $\hat{\cal E'}$,$\hat{\psi'}$ describe 
the same static BR fields as the Ernst potentials $\hat{\cal E}$,$\hat{\psi}$ of the original 
BR solution, but in a different coordinate frame.  
For the special choice $\gamma = (1 + \eta^2)^{-1/2}$ (corresponding 
to $\hat{\cal E}' = -1$), 
the final transformation $\Pi^{-1}$ then leads to the Ernst potentials 
\be
{\cal E}' = \frac{px - iqy - 1}{px - iqy + 1}\,, \qquad \psi' = 0
\ee
(with $p = (1+\eta^2)^{-1/2}$, $q = \eta(1+\eta^2)^{-1/2}$), which are those of
the Kerr solution \cite{kerr}. Summarizing, the Kerr solution has been generated 
from   the Schwarzschild solution by the transformation $\Sigma = \Pi^{-1}{\cal R}\Pi$.
The choice of another value for $\gamma$ would lead after the second step to a 
constant potential $\hat{\cal E}' \neq -1$, resulting in a final electromagnetic 
potential $\psi' \neq 0$. 
So for a generic value of $\gamma$ the transformation $\Sigma = 
\Pi^{-1}{\cal R}\Pi$ generates the Kerr--Newman family of solutions 
from the Schwarzschild solution. 

Another example is that of the Voorhees--Zipoy family of static vacuum
solutions \cite{zip}, depending on a real parameter $\delta$, with the Ernst 
potentials (in prolate spheroidal coordinates)
\be
{\cal E} = \left( \frac{x-1}{x+1} \right)^{\delta} \,, \qquad \psi = 0\,.
\ee
Carrying out the spin--generating transformation $\Sigma$, with $\gamma$ 
chosen such that the resulting solutions are electrically neutral, we have 
obtained in \cite{kerr} a family of rotating solutions
depending continuously on three parameters.  
These solutions, which have a dipole magnetic moment and a quadrupole electric 
moment, are clearly different from the discrete Tomimatsu--Sato \cite{ts}
family of vacuum rotating solutions, which also reduce to Voorhees--Zipoy
solutions in the static limit, but are known only for integer $\delta$. 

Yet another example is the construction of self--gravitating cosmic rings. 
It has been shown \cite{davis} that superconducting closed cosmic strings, 
which would otherwise collapse under the effect of their own tension, 
can be stabilized by rotation. As shown in \cite{lring}, such rotating cosmic 
strings or vortons without thickness can be obtained as self--consistent 
solutions to the Einstein--Maxwell equations. The starting point is the 
charged static ``ring wormhole'' solution \cite{kb}, given in oblate
spheroidal coordinates $(x, y)$ by
\be
{\cal E} = +1\,, \quad \psi = \cot{\sigma}\,, \qquad\sigma(x) = \sigma_0
+ \alpha\arctan\,x\,.
\ee
The corresponding spacetimes have a circular topological defect (cosmic ring) 
at $x = y = 0$. The transformation $\Sigma$ generates from these static 
solutions a family of exact rotating solutions depending on 4 parameters,
which reduce to 3 in the electrically neutral case; these parameters
are further constrained by the condition that the ring $x = y = 0$ rotates
at subluminal velocities. The resulting spacetimes have an ``extra'' point 
at spatial infinity (which however cannot be reached by timelike or null 
geodesics) and may also present (depending on the parameters) an extra ring 
singularity.

In the case of a generic axisymmetric electrostatic solution (${\cal E}$,
$\psi$ real), it is convenient to choose Weyl
coordinates $\rho$, $z$, $\varphi$ such that
\begin{equation}
\omega_i\,dx^i \equiv \omega(\rho,z)\,d\varphi\,,\qquad h_{ij}\,dx^i\,dx^j
= {\rm e}^{2k(\rho,z)}(d\rho^2 + dz^2) + \rho^2\,d\varphi^2\,.
\end{equation}
The first two Ernst equations (\ref{ernst2}) for the transformed BR--like 
Ernst potentials $\hat{{\cal E}}$, $\hat{\psi}$ then take the real form 
\begin{equation}
\nabla(\rho\hat{f}^{-1}\nabla\hat{{\cal E}}) = 0\,,\qquad 
\nabla(\rho\hat{f}^{-1}\nabla\hat{\psi}) = 0\,,
\end{equation}
which imply the existence of dual Ernst potentials $\hat{\cal F}$,
$\hat{\phi}$ such that 
\begin{equation}\label{dual}
{\hat{\cal F}}_{,m}  = \rho\hat{f}^{-1}\,\epsilon_{mn}{\hat{{\cal E}}}_{,n}\,,
\quad {\hat{\phi}}_{,m} = \rho\hat{f}^{-1}\,\epsilon_{mn}{\hat{\psi}}_{,n}
\end{equation}
($m$,$n$ = 1,2, with $x^1=\rho$, $x^2=z$). It may then be shown that the 
transformation ${\cal R}$ with $\gamma = 1$ transforms the potentials 
($\hat{{\cal E}}$, $\hat{\psi}$, ${\rm e}^{\hat{2k}}$) into
\begin{eqnarray}\label{crank}
\hat{{\cal E}}' & = & \hat{{\cal E}} + 2i\Omega\,(z + \hat{\cal F} + 
\hat{\psi}\hat{\phi}) - \Omega^2 (\rho^2/\hat{f} +
\hat{\phi}^2)\,,
\nonumber \\ 
\hat{\psi}' & = & \hat{\psi} + i\Omega\hat{\phi}\,, \qquad {\rm e}^{2\hat{k}'}
= (1 - \Omega^2\rho^2/\hat{f}^2)\,{\rm e}^{2\hat{k}}\,.
\end{eqnarray}
From these one may write down the asymptotically flat potentials
${\cal E}'$, $\psi'$, from which the rotating metric $g_{\mu\nu}'$ and the
rotating electromagnetic potentials $A_{\mu}'$ may be obtained by solving duality
equations. A partial solution of this problem is (again for $\gamma = 1$):
\begin{eqnarray}\label{spin}
f' & = & (|\hat{F}|^2/|\hat{F}'|^2)\lambda\,f\,, \qquad {\rm e}^{2k'} =   
\lambda\,{\rm e}^{2k}\,, \nonumber \\
\omega',_m & = & \Omega^{-1}|\hat{F}'|^2 (\lambda^{-1}),_m 
- (2\rho/\hat{f})\lambda^{-1}\epsilon_{mn}\,
{\rm Im}(\hat{\overline{F}}'\hat{F}',_n)\,,
\end{eqnarray}
with 
\be
|\hat{F}'| \equiv |1 - \hat{{\cal E}}'+ 2\hat{\psi}'|/2  = 1/|F'|\,,
\quad \lambda \equiv (1-\Omega^2\rho^2/\hat{f}^2)\,.
\ee 

From (\ref{spin}) one can derive the main generic properties of 
these rotating Einstein--Maxwell solutions:

1) They are {\it regular} on the axis $\rho = 0$,
\begin{equation}
{\rm e}^{2k'} = 1\,, \qquad \partial_z\omega' = 0\,,
\end{equation}
if the original static fields are regular. 

2) They present {\it ergospheres} $f'(\rho,z) = 0$ for
$\hat{f}(\rho,z) = \pm\Omega\rho$. Near such zeroes of $f'$, 
the rotating metric 
\begin{equation}
ds'^2 \simeq \mp 2\rho\,dt\,d\varphi 
\mp(\hat{F}'^2/\Omega\rho)({\rm e}^{2k}(d\rho^2 + dz^2) +
\rho^2\,d\varphi^2) 
\end{equation}
is non--degenerate.

3) The rotating solutions present {\it horizons} for
\be
f'\e^{-2k'} = \frac{|F'|^2}{|F|^2}\,f\,\e^{-2k} = 0 \,.
\ee
These may occur for either

--a) zeroes of $f$ (horizons of the static solution); an 
illustration of this horizon
conservation is the fact that the transformation $\Sigma$ (with arbitrary
$\Omega$ and $\gamma$) transforms an extreme Reissner--Nordstr{\"o}m black
hole ($M^2 = Q^2$) into an extreme Kerr--Newman black hole ($M'^2 = Q'^2 +
a'^2$); or

--b) poles of $F$ (leading to simple zeroes of $f/|F|^2$); an
example is the inner horizon of the Kerr--Newman solution, which is generated 
from the singularity $r = 0$ of the Schwarzschild solution.

4) The rotating metric (\ref{spin}) may present Kerr--like {\it ring 
singularities}
corresponding to the zeroes of the function $|\hat{F}'|^2(\rho,z)$. In the
plane--symmetric case, these rings are located in the plane $z = 0$ 
(${\rm Im} \hat{F}'= 0$) with radii given by the solutions of the equation
\begin{equation}\label{sing}
2\,{\rm Re}\,\hat{F}' = (1+\hat{\psi})^2 - \hat{f}' =  0 \qquad {\rm for}\,z=0
\end{equation}
($\hat{f}' \equiv \hat{f} - \Omega^2\rho^2/\hat{f}$). Near such rings the 
rotating metric reduces to
\be
ds'^2 \equiv \frac{\hat{f}'_0}{|\hat{F}'|^2}\,(dt - \omega'_0\,d\varphi)^2 -
\frac{|\hat{F}'|^2}{\hat{f}'_0}\,(\e^{2\hat{k}'_0}(d\rho^2 + dz^2) 
+ \rho_0^2\,d\varphi^2)\,,
\ee
with $\hat{f}'_0 \ge 0$ from Eq. (\ref{sing}), so that the rotating ring singularity 
is timelike if $\hat{f}'_0 \neq 0$ (the Kerr ring corresponds to the very 
special case in which $\hat{f}'_0$ also vanishes on the ring). In the case of the
solutions studied in \cite{lring}, these unwanted ring singularities may
be avoided by suitably choosing the parameters of the static solution
and/or the parameter $\Omega$.

There are surely other interesting applications of this solution--generating
technique besides those mentioned here. At a more fundamental level, there
remain a number of open questions.

A first question is that of the generality of the transformation $\Sigma$.
For instance, how many inequivalent rotating solutions can be generated 
from a given static solution by transformations ${\cal
U}\Sigma{\cal U}^{-1}$, with ${\cal U} \in {\rm SU(2,1)}$? For those
transformations ${\cal U}$ which conserve staticity, the resulting solutions
are found to belong the same family (with different parameter values), 
however the case of more general transformations ${\cal U}$ has not yet been 
elucidated. A related
question is that of the precise connection of the direct spin--generating
method presented here with other spin--generating techniques. This method 
could also in principle be generalized to other gravitating field theories
with a highly symmetric stationary target space. Two cases are presently 
under investigation, those of five--dimensional Kaluza--Klein theory, and of
dilaton--axion gravity with one Abelian gauge field.

\end{document}